\def\aa{A\&A}               
\def\aas{A\&AS}             
\def\anj{AJ}                
\def\apj{ApJ}               
\def\mn{MNRAS}              
\def\nat{Nature}            
\def\pasp{PASP}             
\def\rev{ARA\&A}            
\def\apss{Ap\&SS}           

\documentstyle{aa}

\begin{document}

\title{A new sample of large angular size radio galaxies}
\subtitle{III. Statistics and evolution of the grown population}

\author{L. Lara\inst{1,2} 	\and 
G. Giovannini\inst{3,4} 	\and
W.D. Cotton\inst{5}     	\and
L. Feretti\inst{3}      	\and
J.M. Marcaide\inst{6}   	\and
I. M\'arquez\inst{2}    	\and
T. Venturi\inst{3}
}

\offprints{L. Lara \email{lucas@ugr.es}}

\institute{Dpto. F\'{\i}sica Te\'orica y del Cosmos, Universidad de
Granada, Avda. Fuentenueva s/n, 18071 Granada (Spain)
\and
Instituto de Astrof\'{\i}sica de Andaluc\'{\i}a (CSIC),
Apdo. 3004, 18080 Granada (Spain)
\and
Istituto di Radioastronomia (CNR), via P. Gobetti 101, 40129 Bologna (Italy)
\and
Dipartamento di Astronomia, Universit\'a di Bologna, via Ranzani 1,
40127 Bologna (Italy)
\and 
National Radio Astronomy Observatory, 520 Edgemont Road, Charlottesville, 
VA 22903-2475 (USA)
\and
Departimento de Astronom\'{\i}a, Universitat de Val\`encia, 46100 Burjassot - 
Spain
} 
\date{Received / Accepted}

\authorrunning{L. Lara et al.}
\titlerunning{A new sample of large angular size radio galaxies. III.}

\abstract{
We present in this paper a detailed study of a new sample of large
angular size FR I and FR II radio galaxies and compare the
properties of the two classes. As expected, a pure morphology based
distinction of FR Is and FR IIs corresponds to a break in total radio
power. The radio cores in FR Is are also weaker than in FR IIs,
although there is not a well defined break power. 
We find that asymmetry in the structure of the sample members must be 
the consequence of anisotropies in the medium where the lobes expand,
with orientation playing a minor role. Moreover, literature data and
our observations at kiloparsec scales suggest that the large
differences between the structures of FR I and FR II radio galaxies
must arise from the poorly known central kiloparsec region of their
host galaxies. We analyze the sub-sample of giant radio galaxies, and
do not find evidence that these large objects require higher core
powers. Our results are consistent with giant radio galaxies being the
older population of normal FR I and FR II objects evolving in low
density environments. Comparing results from our sample with
predictions from the radio luminosity function we find no evidence of
a possible FR II to FR I evolution. Moreover, we conclude that at
$z\sim 0.1$, one out of four FR II radio sources has a linear size
above 500 kpc, thus being in an advanced stage of
evolution (for example, older than $\sim 10$ Myr assuming a jet-head 
velocity of 0.1c). Radio activity seems to be a short-lived process
in active galaxies, although in some cases recurrent: five objects in
our sample present signs of reactivation in their radio structures.
\keywords{Galaxies: active -- Galaxies: nuclei -- Galaxies: jets  -- 
Radio continuum: galaxies}
}

\maketitle

\section{Introduction}

Radio galaxies, with linear sizes reaching up to several megaparsecs,
are possibly the largest individual objects in the Universe.  It is
widely accepted that they originate from highly energetic non-thermal
processes occurring in the nucleus of the so-called active galaxies
(Blandford \& Rees \cite{blandford}; Rees \cite{rees1}). According to
the standard model of active galactic nuclei (AGNs), a super-massive
black hole with a mass between $10^6$ and $10^9 M\odot$, resides
in the center of the active galaxy, powered by an accretion disk
surrounded by a torus formed by gas and dust. In about 10\% of these
AGNs, there is intense synchrotron radio emission produced in a
bipolar outflow of relativistic particles expelled perpendicularly to
the plane of the disk and extending to distances reaching the
megaparsec scales.  The reason why an AGN presents or not powerful
radio emission is matter of strong debate.  While there is increasing
evidence about the existence of super-massive black holes in the center
of AGNs, and even at the nuclei of non active galaxies (Macchetto
\cite{macchetto}; Kormendy \& Gebhardt \cite{kormendy}), it has been argued
that the presence or not of intense radio emission might be due to
the rotation velocity of the black hole (e.g. Wilson \& Colbert
\cite{wilson}; Cavaliere \& D'Elia \cite{cavaliere}), or to its total
mass and the efficiency of accretion (McLure \& Dunlop \cite{mclure};
Dunlop et al. \cite{dunlop2}).

Considering a natural evolutionary sequence of radio galaxies, jets
emanating from the center of activity start boring their way through
the interstellar medium first, reaching the galactic halo and in some
large cases the intergalactic medium. Finally, the lobes of radio
galaxies which have ceased their central activity expand and disappear
in the external medium. Radio sources representing these different
phases of evolution are currently known adding support to this
scenario: compact symmetric objects
(CSOs; Wilkinson et al. \cite{wilkinson}) are thought to be young
radio galaxies (e.g. Owsianik \& Conway \cite{owsianik}), while the
giant radio galaxies (GRGs; defined as those with a projected linear
size\footnote{We assume that H$_{0}=50$km s$^{-1}$ Mpc$^{-1}$ and
q$_{0}=0.5$} $\ge$ 1 Mpc) are probably old objects at the latter
stages of evolution (Ishwara-Chandra \& Saikia \cite{ishwara}). Relic
radio sources found in clusters of galaxies might correspond to the
last detectable emission from ``dead'' radio galaxies (e.g. Komissarov \&
Gubanov \cite{komissarov}; Slee et al. \cite{slee}).

However, the degree of influence of parameters other
than the age (e.g. source power, conditions of the external medium) in
the evolution of radio galaxies is not clear . For example, although there is
observational evidence supporting the young source scenario for CSOs,
it has also been argued that CSOs are short lived objects which never
reach the size of their big relatives (Readhead et
al. \cite{readhead}). At the other extreme, GRGs could be the result
of normal radio galaxies expanding in very low density environments
permitting them to reach their overwhelming sizes.  But they could
also result from very powerful core activity, or both conditions must
apply for a radio galaxy to become a giant. Complicating the
previously outlined evolutionary sequence, some radio galaxies seem to
wake up after a dormant phase of absence or much lower activity
(e.g. Lara et al. \cite{lara}).  Moreover, the presence of
super-massive objects in many non--active galaxies argue in
favor of activity as a short transition period in most, if not all,
(elliptical) galaxies, and that the ``menace" for future activity is
present at the center of every galaxy.
 
This paper is the last of a series of three devoted to the study of a
sample of large angular size radio galaxies which try to address
some of these open questions. Definition of the sample and radio maps
in one side, and images and spectroscopic data on the other, were
presented by Lara et al. (\cite{paperI}, hereafter Paper I) and
(\cite{paperII}, Paper II), respectively. The sample, covering a sky
area of 0.842 steradians\footnote{Papers I and II erroneously mention
a sky area of $\pi$ steradians. However, results and discussion in
those papers are not affected by 
this value.} and spectroscopically complete at the 80\%,
consists of 84 radio galaxies selected from the NRAO\footnote{National
Radio Astronomy Observatory} VLA\footnote{Very Large Array, operated
by the NRAO} Sky Survey (NVSS; Condon et al. \cite{nvss}) under the
following selection criteria (see Paper I for details): declination
above $+60^{\circ}$, total 
flux density at 1.4 GHz greater than 100 mJy and angular size larger
than 4$\arcmin$. All sources in the sample have redshift below
0.75. In this paper we present the general results of the study of the
sample, with distinction between Fanaroff-Riley type I (FR I; Fanaroff
\& Riley \cite{fanaroff}) and II (FR II) radio galaxies. Special
attention is devoted to show the properties of GRGs, of which this
sample contains 37 members. 

\section{General properties}

\subsection{FR I and FR II radio galaxies}
\label{fr12}

Fanaroff \& Riley (\cite{fanaroff}) distinguished two classes of radio
galaxies, with clearly distinct properties: a first group (FR Is) of
low power sources with radio structures dominated by the emission from
the core and the jets, with lobes gradually decreasing in
brightness with distance, and a second group (FR IIs) of higher power
sources, with structures dominated by prominent edge-brightened radio
lobes with hotspots at the position where the jet impinges on the
external medium. From a pure morphological distinction of the radio
structure of the members of our sample, regardless of the source radio
or optical luminosity, we find 31 sources (37\%) which
correspond to the FR I type, and 46 sources (55\%) which correspond to
the FR II type, of which three of them are classified as quasars. 
Six objects (7\%) are difficult to classify in any of the two groups,
showing simultaneously properties of FR I and FR II radio galaxies. We
put them in an intermediate group, labeled as I/II (see Gopal-Krishna
\& Wiita \cite{gopal2}). Moreover, there is \object{J1015+683}  which results
from the superposition in projection of a possibly FR I and a FR II radio
sources (see Fig.~2 in Paper II).

\begin{figure}
\vspace{12cm}
\includegraphics{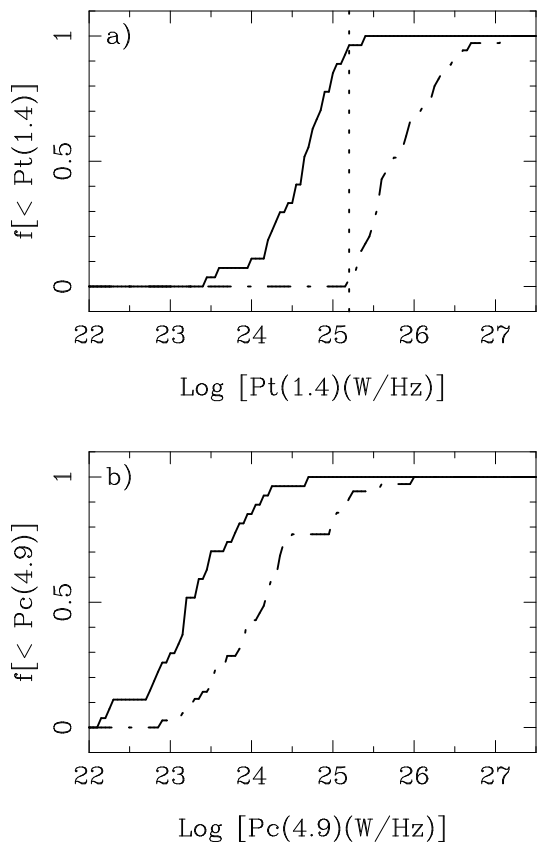} 
\caption{{\bf a} Integrated fraction of sources with total radio power
at 1.4 GHz below $P_t$. Continuous and dotted-dashed lines stand for
FR I and FR II type radio galaxies, respectively. The vertical dotted
line represents the break power between FR I and FR II radio
galaxies. {\bf b} Integrated fraction of sources with core radio power
at 4.9 GHz below $P_c$.} 
\label{lum_funct}
\end{figure}

In Fig.~\ref{lum_funct}a we represent the fraction of sources
with reliably measured redshift and total radio power below
$P_t$(1.4). We note how the morphologically based distinction between
FR Is and FR IIs corresponds to a neat break in source total
power: 96\% of the FR I radio sources have $\log P_t(1.4) \le 25.2$
(W/Hz), while 97\% of the FR IIs have higher radio power. This
sharp break in radio power contrasts with the result by Ledlow \& Owen
(\cite{ledlow}) who find for a large sample of radio galaxies that the
FR I to FR II division is a function of both, radio power and optical
luminosity of the host galaxy.
FR I/II radio sources, not represented in Fig.~\ref{lum_funct}, fall in the
region of transition between FR Is and FR IIs, with a mean $\log
P_t(1.4) = 25.4$ (W/Hz). 

Fig.~\ref{lum_funct}b represents the same
concept as Fig.~\ref{lum_funct}a, but considering the core radio power
instead of the total radio power. We find that FR Is have in general
weaker cores than FR IIs, but a sharp distinction in the core power of
the two classes cannot be established.

\begin{figure}
\vspace{9.5cm}
\includegraphics{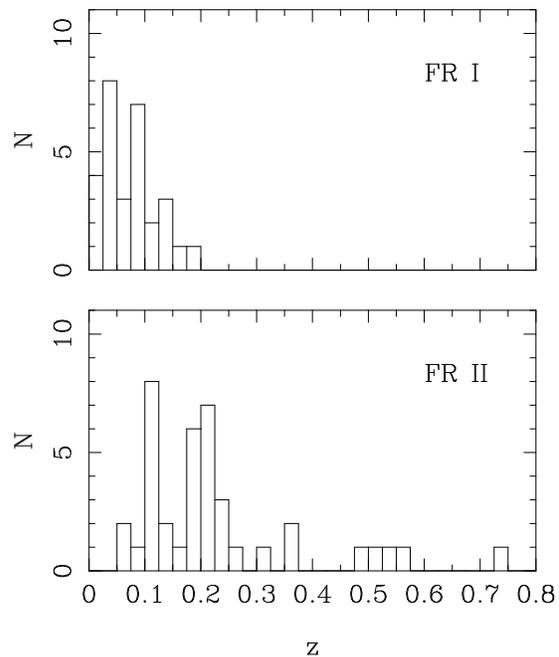} 
\caption{Distribution of redshift in our sample for FR I (top) and FR
II (bottom) radio galaxies. The redshift bin is 0.025} 
\label{z_histo}
\end{figure}

Regarding the redshift distribution of the sample members, the
different power and radio structure of FR I and FR II type radio
sources introduce a clear selection effect. Since FR Is have low power
and are diffuse and extended, surface brightness sensitivity
limitations will hinder in some cases the detection of their most
extended emission, preventing many FR Is to reach the $4'$ limiting
size of the sample\footnote{We refer to Paper I for the source size
definition}. FR IIs, on the contrary, end in bright and compact
hotspots, so their total size can be reliably measured. Another
consequence of this fact is that FR Is with sizes larger than 4$'$ are
difficult to detect at large cosmological distances, since that would
imply intrinsically bright
extended emission, which is not usual in this type of sources. In
consequence, we expect most FR Is to be at low redshift. On the other
hand, for FR IIs we do not have any {\em a priori} selection effect,
so in principle they can be found at low and at high redshifts. In
Fig.~\ref{z_histo} we show the distribution of redshift for the FR I
and the FR II sources in our sample. As expected, we find indeed that
all FR Is have $ z \le 0.2$, with a median value $\langle z_{FR\,I}
\rangle = 0.07 $. On the other hand, the distribution with redshift of
FR II radio galaxies is much broader, with a median value $\langle
z_{FR\,II} \rangle = 0.20$. In Section~\ref{lumf} we discuss the
distribution in redshift under the perspective of the radio luminosity
function.

\subsection{Radio source asymmetry}
\label{asym}

We define the arm-length ratio ($r$) of a radio source as the ratio
between the length of its shorter to its longer arm. Arm-lengths
were measured along the spine of the sources using our VLA maps at 1.4
GHz, or the NVSS maps when there was extended emission resolved out in
our observations (see Paper I). The arm-length ratios of the members
of the sample are displayed in Table~\ref{armlratio}. Errors in the
determination of $r$ are in the range of 2\% to 5\%, with FR I sources
having larger errors than FR IIs. This is mostly due to the
difficulties associated with the determination of FR I sizes through
isophotal measurements (sensitivity dependent), in contrast with the
metric and well defined sizes of FR II radio galaxies. But given the
high sensitivity of our observations and of the NVSS, we have
confidence that arm-length ratio measurements are reliable in most
cases. Possible exceptions will not alter the statistics significantly.

\begin{figure}
\vspace{9.5cm}
\includegraphics{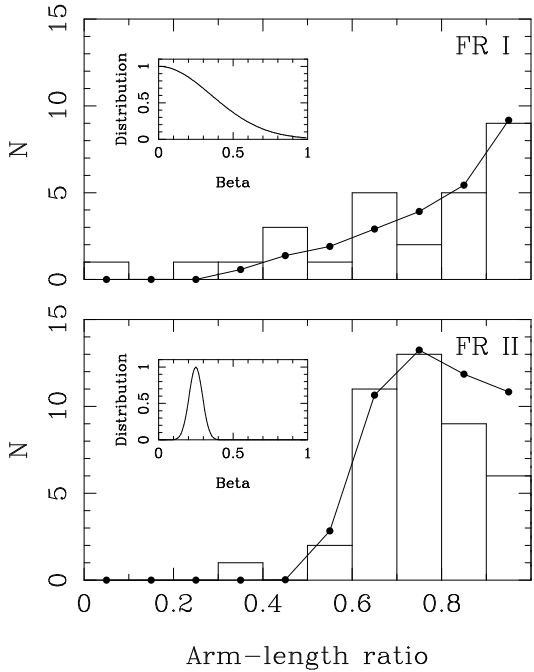} 
\caption{Distribution of the arm-length ratio for FR I (top) and FR II 
(bottom) radio galaxies. The filled circles and lines represent the
prediction of an orientation-based model considering the jet-head
advance velocity Gaussian distribution plotted in the small frame.} 
\label{arm}
\end{figure}

The distribution of parameter $r$ is plotted in
Fig.~\ref{arm}. We find a clear difference between the distribution
for FR I and FR II sources. While the distribution of the former
gradually increases towards symmetry, with 50\% of the sources
presenting $r\ge 0.8$, the distribution of the latter appears rather
concentrated around a mean value of $0.78\pm 0.14$. 

\begin{table*}[t]
\caption[]{Arm-length ratio ($r$) of the sample members}
\label{armlratio}
\begin{scriptsize}
\begin{tabular}{llllllll}
\hline \hline
Name    & ~~r & Name & ~~r & Name & ~~r & Name & ~~r \\
\hline	 
J0109+731 & 0.58 & J0807+740 & 0.91 & J1504+689 & 0.61 & J1918+742 & 0.37 \\
J0153+712 & 0.09 & J0819+756 & 0.70 & J1523+636 & 0.70 & J1951+706 & 0.79 \\
J0317+769 & 0.29 & J0825+693 & 0.89 & J1530+824 & 0.62 & J2016+608 & 0.60 \\
J0318+684 & 0.73 & J0828+632 & 0.57 & J1536+843 & 0.90 & J2035+680 & 0.66 \\
J0342+636 & 0.65 & J0856+663 & 0.71 & J1557+706 & ~--  & J2042+751 & 0.95 \\
J0430+773 & 0.88 & J0926+653 & 0.93 & J1632+825 & ~--  & J2059+627 & 0.90 \\
J0455+603 & 0.88 & J0926+610 & 0.66 & J1650+815 & 0.60 & J2103+649 & 0.70 \\
J0502+670 & 0.95 & J0939+740 & 0.91 & J1732+714 & 0.98 & J2111+630 & 0.55 \\
J0508+609 & 0.74 & J0949+732 & ~--  & J1733+707 & 0.92 & J2114+820 & 0.98 \\
J0519+702 & 0.96 & J1015+683 & ~--  & J1743+712 & 0.61 & J2128+603 & 0.65 \\
J0525+718 & 0.95 & J1036+677 & 0.78 & J1745+712 & 0.87 & J2138+831 & 0.47 \\
J0531+677 & 0.34 & J1124+749 & 0.58 & J1751+680 & 0.63 & J2145+819 & 0.98 \\
J0546+633 & 0.47 & J1137+613 & 0.80 & J1754+626 & 0.42 & J2157+664 & 0.09 \\
J0559+607 & 0.82 & J1211+743 & 0.67 & J1800+717 & 0.74 & J2204+783 & 0.79 \\
J0607+612 & 0.88 & J1216+674 & 0.73 & J1835+665 & 1.00 & J2209+727 & 0.92 \\
J0624+630 & 0.78 & J1220+636 & 0.75 & J1835+620 & 1.00 & J2242+622 & 0.83 \\
J0633+721 & 0.67 & J1247+673 & 1.00 & J1844+653 & 0.93 & J2247+633 & 0.89 \\
J0654+733 & 0.77 & J1251+756 & 0.79 & J1845+818 & 0.69 & J2250+729 & 0.84 \\
J0750+656 & 0.91 & J1251+787 & 0.47 & J1847+707 & 0.87 & J2255+645 & 0.92 \\
J0757+826 & 0.66 & J1313+696 & 0.77 & J1850+645 & 0.81 & J2307+640 & 0.82 \\
J0803+669 & 0.70 & J1410+633 & 0.78 & J1853+800 & 0.71 & J2340+621 & 1.00 \\
\hline 
\end{tabular}
\end{scriptsize}
\end{table*}

Apparent asymmetry is common in radio galaxies, and in principle three
possible reasons could be envisaged to explain it. {\bf First}, plasma
ejection from the active nucleus could have different properties for
each one of the jets (confinement, velocity, etc.) producing a
different radio source morphology at each side, and in particular, a
different size of each arm in the radio galaxy. {\bf Second}, the different
arm size could be due to an orientation effect. If the source is not
lying on the plane of the sky, the differing light travel times to the
observer for each arm will produce different sizes. And {\bf third}, the
external medium through which the jets propagate might not be
isotropic, presenting different resistance to the propagation of the
jets, and thus producing different size arms.

Although these three effects could act in a combined manner, we can
try to constrain the possible reasons of asymmetry in our sample.  
First we note that the wealth of observations of the inner jets in radio
galaxies and quasars do not show evidence for intrinsic
differences in the properties of jets and counter-jets. The observed
differences are readily explained through the effects of relativistic
aberration in symmetric, anti-parallel jets (e.g. Giovannini et
al. \cite{giov01}; Laing \& Bridle \cite{laing1}). In the 
following, we thus concentrate in distinguishing between the external
medium or source orientation as the main reason of asymmetry.  

We have considered a randomly oriented sample
($P(\theta)\propto\sin\theta$; $P$ standing for probability) of
intrinsically symmetric radio galaxies with a Gaussian advance
velocity distribution of the jet head. Under these assumptions, we are
able to obtain reasonable approximations to the shape of the observed
arm-length ratio distribution solely on the basis of orientation
effects for both, FR I and FR II radio sources, using different jet
velocity distributions for each type of radio galaxy (see
Fig.~\ref{arm}). But what is relevant here, our model requires too
high (and thus unrealistic) jet-head advance velocities for FR Is and
FR IIs. For example, a velocity distribution narrowly centered around
0.25c is required to explain the arm-length ratio distribution of FR
II radio galaxies, while most models and observational data suggest
much lower expansion velocities (e.g. Longair \& Riley \cite{longair};
Alexander \& Leahy \cite{alexander}; Scheuer
\cite{scheuer2}). Moreover, even with unacceptable jet--head  
velocities, the model still predicts more symmetric sources than
observed which might reflect the influence of the external medium not
accounted for. 

As a second argument, we find an strikingly large fraction
($\sim 75$\%) of FR II radio galaxies with a stronger lobe on the
shorter arm of the radio structure. A similar behavior is also found
by Machalski et al. (\cite{machalski}) in a sample of GRGs. To compute
this percentage we have excluded quasi-symmetric sources (with $0.9\le
r \le 1$) to avoid confusion. We note that in an orientation based
asymmetry, it is expected that the shorter arm corresponds to the
receding jet and if the emission from the lobes is at least moderately
beamed (e.g. Georganopoulos \& Kazanas \cite{georga}), it is expected
that the receding lobe presents 
weaker emission than the approaching one. Therefore we suggest that
orientation cannot be the main reason of asymmetry.

The asymmetry could be explained if the external medium were not
isotropic. In that case the shorter lobe would be the one finding
stronger resistance to its expansion in the external medium, which
shows up as a higher surface brightness. Supporting this idea, the sources
for which we have polarization measurements present also a tendency of
stronger polarized emission in the shorter and stronger lobe,
consistent with a compression of the magnetic field against the
external medium, but this fact needs to be confirmed through more detailed
observations. Moreover, the anisotropies in the external medium most plausibly
explain the existence of several sources with pairs of jets showing
markedly different properties or with a hybrid FR I/II morphology. The
most dramatic case is \object{J2157+664} (see Paper I), with $r=0.09$.

In conclusion, with the necessary caution required due to the
uncertainties in the arm-length ratio determinations in FR Is, we
obtain {\em i)} from the orientation based asymmetry model and
{\em ii)} from the fact that many sources show the stronger lobe on the
shorter arm, that the external medium must be the dominant effect in
most of the asymmetric sources of our sample. However, we cannot
exclude that orientation effects might play their role to explain
certain degree of asymmetry in FR Is and FR IIs. Moreover, we do not
find a clear dependence of the arm-length ratio of FR IIs with the
source size. 

\subsection{Radio core properties}
\label{pcore}

A compact radio core is detected at 4.9 GHz in 100\% of the sources in the
sample. The core spectral index\footnote{The spectral index $\alpha$
is defined so that the flux density $S\propto \nu^{\alpha}$}
distribution of the whole sample has 
$\langle \alpha \rangle = -0.10 \pm 0.47 $. If we consider separately
the FR I and the FR II type radio galaxies, we find $\langle \alpha_{FR\,I}
\rangle = -0.24 \pm 0.52 $, while $\langle \alpha_{FR\,II} \rangle = 0.07
\pm 0.41 $. We ascribe the different mean
values of the two families of radio galaxies to the fact that the
radio core in FR Is cannot be isolated from the jets as neatly as
in FR IIs. In consequence, the core spectral index in FR Is suffers
from more contamination from the steeper jet emission than FR IIs.
This result is consistent with that found in the B2 sample (de Ruiter
et al. \cite{deruiter}).

\begin{figure}
\vspace{6cm}
\includegraphics{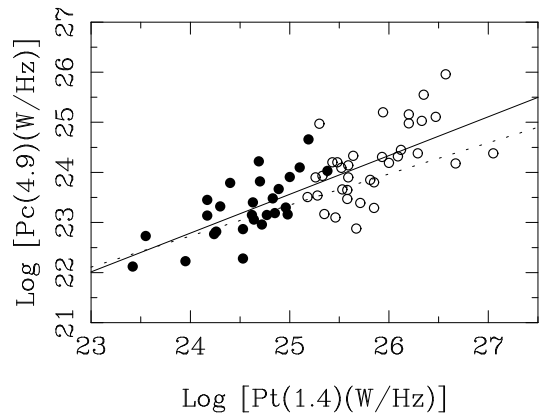}
\caption{Total power at 1.4 GHz vs core power at 4.9 GHz. The
continuous line represents a least squares fit to our data. The dotted
line represents the correlation $\log P_c$(4.9\,GHz)$ = (0.62\pm 0.04) \log
P_t$(0.408\,GHz)$ + (7.6\pm 1.1)$ by Giovannini et al. (\cite{giov01})
transformed to $P_t$(1.4\,GHz), assuming an spectral index
$\alpha=-0.75$. Filled and empty circles correspond to FR I and FR II
type radio galaxies, respectively.} 
\label{Pt1_Pc5}
\end{figure}

In Fig.~\ref{Pt1_Pc5} we display the core radio power at 4.9 GHz as a
function of the total radio power at 1.4 GHz. We find a correlation
between these two parameters of the form
\[
\log P_c(4.9) = (0.77\pm 0.08) \log P_t(1.4) + (4.2\pm 2.1).
\]

This correlation is steeper than that of Giovannini et
al. (\cite{giov01}; dotted line in Fig.~\ref{Pt1_Pc5}, where we have
transformed Giovannini's correlation from $P_t$(408 MHz) to $P_t$(1.4
GHz) assuming a spectral index $\alpha =-0.75$), but it is still
consistent with it. As discussed by several authors (e.g. Giovannini
et al. \cite{giov94}) the prominence of the core with respect to the
total radio emission can be considered as an indicator of orientation
of the nuclear jet, and, more generally, of the whole radio source
with respect to the observer. 

\begin{figure}
\vspace{8cm}
\includegraphics{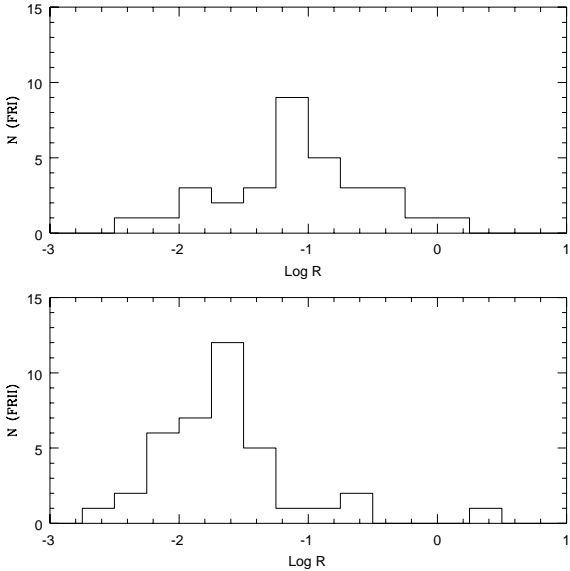}
\caption{Distribution of the orientation indicator $R$ (see text) in
logarithmic scales, for FR Is (top) and FR IIs (bottom) radio galaxies}
\label{luigi1}
\end{figure}

Morganti et al. (\cite{morganti}) define as an orientation parameter
the ratio $R = S_{core} /(S_{tot} - S_{core})$, i.e. the ratio between
the flux density of the core and the flux density of the extended
structure, both at 1.4 GHz. We have derived this parameter for our
sample, and show in Fig.~\ref{luigi1} the histogram of $\log R$ for
the FR I and the FR II classes. We find that the FR Is are
characterized by more dominant cores, and the median values are $\log
R = -1.09$ for the FR Is and $\log R = -1.64$ for the FR-IIs. The
probability that the two histograms are drawn from the same population
is less than 0.1\%, checked with the Kolmogorov-Smirnov test. The
different behavior of the two classes is in the same direction as the
weak difference detected by Morganti et al. (\cite{morganti}). It
could reflect the fact that the core flux density in  FR Is is more
likely to be affected by the jet emission than in FR IIs (the same
reason argued to explain the spectral index difference), and also
that the core radio power in radio galaxies does not increase linearly
with the total radio power in agreement with the slope of the
existing correlation between the core luminosity $P_c$ and the total
luminosity $P_t$.

A more reliable parameter to define the core prominence is then
obtained using the core emission at 4.9 GHz. Moreover, given the
$P_c$ vs $P_t$ correlation, expected if sources are in energy
equipartition conditions (see Giovannini et al.~\cite{giov01}), it is
useful to use as orientation indicator a parameter, $P_{CN}$, defined
as the ratio between the observed core luminosity ($P_c$) and the core
luminosity expected from the $P_c$ vs $P_t$ correlation
($P_{cm}$). The median value of the core luminosity $P_{cm}$ at 4.9
GHz as a function of the total power $P_t$ at 1.4 GHz is given for a
large sample of radio galaxies and quasars by the relation $\log
P_{cm} = 0.62 \log P_t + 8.0 $ obtained by Giovannini et
al. (\cite{giov01}) and scaled here to the total power at 1.4 GHz
using a spectral index $\alpha = -0.75$. We then calculate $P_{CN} =
P_c /P_{cm}$. Considering that in a randomly oriented sample as that
of Giovannini, the average orientation angle with respect to the
observer corresponds to $60^{\circ}$, in presence of relativistic
beaming effects we expect that sources with the angle between jet
axis and line of sight $<60^{\circ}$ or $>60^{\circ}$ show $P_{CN} >1$
or $<1$, respectively. 

\begin{figure}
\vspace{8cm}
\includegraphics{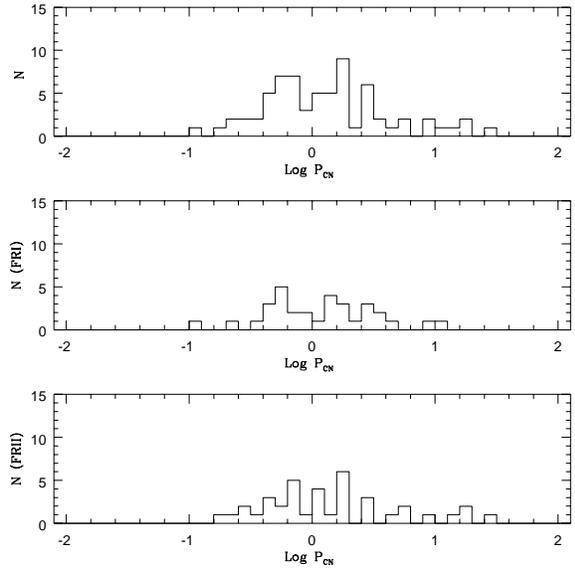}
\caption{Distribution of the orientation indicator $P_{CN}$ (see text)
in logarithmic scales, for the whole sample (top), FR Is (middle) and
FR IIs (bottom) radio galaxies}
\label{luigi2}
\end{figure}

\begin{figure}
\vspace{8cm}
\includegraphics{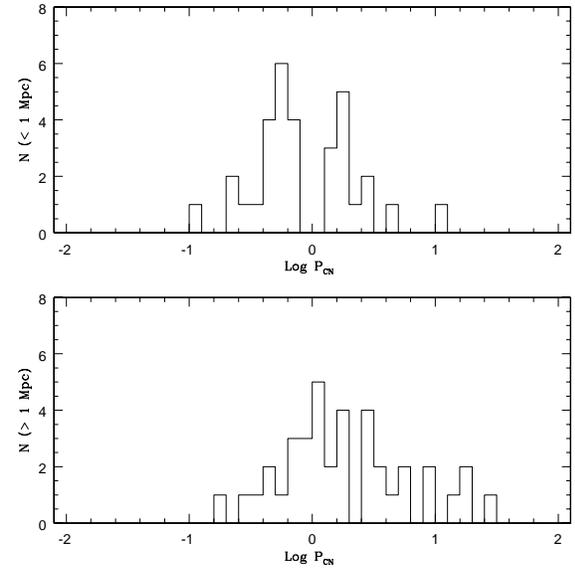}
\caption{Distribution of parameter $P_{CN}$ in logarithmic scales for
sources with projected linear sizes smaller (top) and
larger (bottom) than 1 Mpc}
\label{luigix}
\end{figure}

The distribution of $P_{CN}$ for the whole sample is given in the top
panel of Fig.~\ref{luigi2}, whereas the middle and bottom panels show
the distributions for the FR I and FR II classes, respectively. 
We note that, although Figs.~\ref{luigi2} and \ref{luigix}
display the distribution of $\log P_{CN}$, the considerations in the
following text refer to the values of $P_{CN}$ (not to their logarithm).
We find a large number of sources with $P_{CN}>1$,
i.e. with  a core power larger than expected from their total power
(the median of $P_{CN}$ for the whole sample is 1.22). The same
behavior is found for the FR I and FR II classes, with no
statistically significant difference between the two distributions, on
a Kolmogorov-Smirnov test (medians of the FR I and FR II sources are
1.25 and 1.21, respectively). This indicates that the sample under
study contains an excess of sources which, according to the standard
interpretation, would be oriented at angles smaller than $60^{\circ}$.

\begin{figure}
\vspace{8cm}
\includegraphics{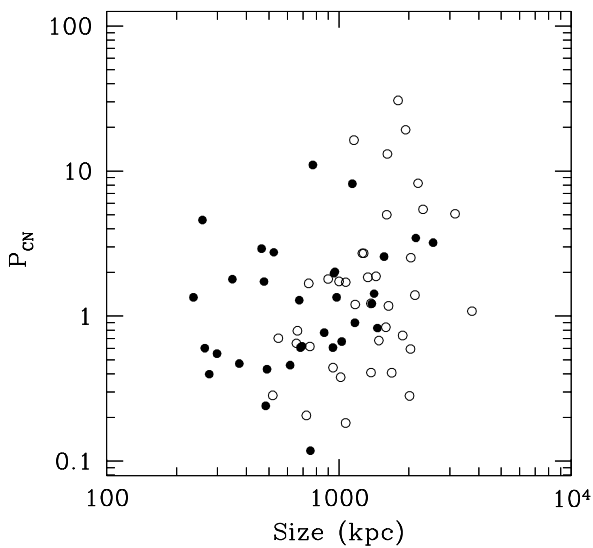}
\caption{Orientation indicator $P_{CN}$ as a function of linear
size. Symbols represent the same as in Fig.~\ref{Pt1_Pc5}}
\label{luigi3}
\end{figure}

This result is quite unexpected, since sources with large apparent
linear sizes are expected to be poorly affected by projection effects,
i.e. preferentially oriented at large angles to the line of sight. In
addition, we would also expect that the sources with the larger
projected linear size are those with a $P_{CN}$ smaller than 1.
This is contrary to the observations.
In fact, by analyzing the distribution of $P_{CN}$ in the sources with
projected linear size smaller and larger than 1 Mpc, we find that the
latter have an excess of values of $P_{CN}>1$ (see Fig.~\ref{luigix}). 
The same trend is deduced by the plot of the parameter $P_{CN}$ 
versus the source linear size (see Fig.~\ref{luigi3}, where FR I and
FR II are indicated as full and empty circles, respectively).
The statistical difference between the two distributions is at the
confidence level of about 2\% on a  
Kolmogorov-Smirnov test. We note that the major source of
uncertainty in the estimation of $P_{CN}$ is related to the radio
galaxy core variability. Assuming a core variability of a factor 2 (much larger
than the core variability observed in most radio sources) we have an
uncertainty of 0.3 in $\log P_c$. However this is the uncertainty
related to a single object; assuming a random core variability we
estimate the possible uncertainty for Fig.~\ref{luigi2} and
Fig.~\ref{luigix} to be about 0.1. 

The difference in the distribution of $P_{CN}$ between sources of
size smaller and larger of 1 Mpc is at a marginal level of statistical
significance. However, it is interesting to note that it is in line
with the predictions of the the evolutionary effects detected by 
Ishwara-Chandra \& Saikia (\cite{ishwara}) and by Schoenmakers et
al. (\cite{arno2}), and discussed in Sect.~\ref{power}. According to
these authors, the total luminosity of a radio source decreases as it
expands to very large dimensions.  As a consequence, its nucleus would
appear more prominent than expected from pure orientation effects.  
From our data, there is indication that the $P_c$ vs $P_t$ correlation is
in support of a luminosity evolution in giant radio sources.  We can
roughly estimate that an increase of 10 to 100 times in the total 
power of GRGs would lead to values of $P_{CN} \sim 1$ for these
sources. Note that a two order of magnitude increase in the total 
power of GRGs would also be consistent with the power-size diagram 
in Fig.~\ref{pd}.

\begin{figure}
\vspace{6cm}
\includegraphics{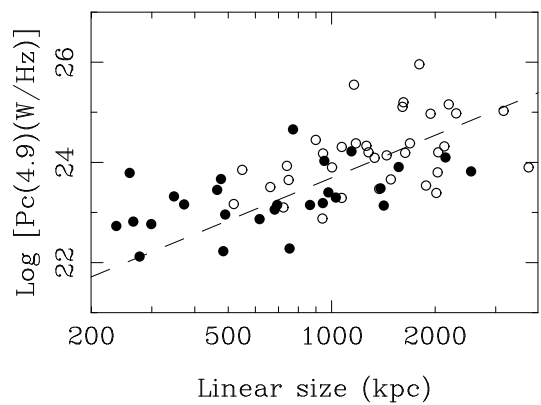} 
\caption{Apparent correlation between the core radio power and the
source linear size, induced by the sample biases. The dashed line
represents the influence of the selection criteria and of the
$P_c$(4.9) vs $P_t$(1.4) correlation in this plot. Symbols represent
the same as in Fig.~\ref{Pt1_Pc5}.} 
\label{D_Pc5}
\end{figure}

We finally note that apparent correlations of the core radio power
with the source redshift and linear size are entirely due to biases
introduced in the sample selection through the flux density and the
angular size limits, and to the correlation between the total and the
core radio power (Paper I). This bias arises from the need for larger
sources to have more flux density to exceed the minimum surface
brightness limit of the NVSS. As an example, we show in Fig.~\ref{D_Pc5}
the core power as a function of the source linear size, which could be
interpreted as sources being intrinsically larger due to an unusually
higher core power (Gopal-Krishna et al. \cite{gopal1}). However, the
dashed line represents how the limits in flux density and angular size
translate into this plot, taking also into account the total power vs
core power correlation mentioned above. We note that it is fully
consistent with the observed trend, so that it is possible to conclude
that the source size is unrelated to the core power, in agreement
with studies on GRGs (Ishwara-Chandra \& Saikia \cite{ishwara}). It
could also be argued that the core power vs source size is a
physically meaningful correlation, and that from the biases introduced
through the selection criteria we could derive the total vs core power
correlation. To isolate the problem in our sample is not easy since we
have both, flux density and angular size limits in our sample
definition, but we note that biases induced correlations depend
strongly on the adopted selection criteria.  However, consistent
correlations of the total vs core power have been obtained with
well defined complete samples including compact
and giant radio sources as well as quasars and radio galaxies (see
e.g. Giovannini et al. \cite{giov88}, \cite{giov01}; de Ruiter et
al. \cite{deruiter}).  

\subsection{The orientation of the sample members}


One of the aims of this work was to select a sample of
radio galaxies with their jets oriented near the plane of the sky for
a subsequent study of the parsec scale properties of these jets. 
Although reliable limits to the orientation of the members of
our sample with respect to the observer cannot be derived from current
data, two lines of argumentation favor the idea that these
sources have moderately large angles of orientation. First, statistics
of radio quasars: we find in our sample 3 objects out of 46 with FR II
radio structure which are classified as quasars (Paper II). If we
consider that the probability of finding an object with a certain angle
$\theta$ to the line of sight is $P(\theta) \sim \sin\theta$, we can
estimate the expected number of quasars in a randomly oriented sample
if quasars are assumed to have orientation angles below $38^{\circ}$
(Urry \& Padovani \cite{urry}). This number is $\sim 10$ for a sample
of 46 FR II radio galaxies, more than three times larger than found in
our sample, which means that most objects must have orientation angles
well above $38^{\circ}$. And second, source intrinsic linear sizes:
with a mean projected linear size of 1.02 Mpc in our sample,
orientation angles below $20^{\circ}$ can, with a high degree of
confidence, be discarded for most objects in order not to have
intrinsically too large radio galaxies, which are not observed in
other samples. 

However, we have found (Section~\ref{pcore}) that the 
correlation between the core power and the total radio power is
consistent with that derived by Giovannini et al. (\cite{giov01}) for a
sample of randomly oriented radio galaxies. The distribution of the
orientation indicator $P_{CN}$ is consistent with a large number of
sources with orientation angles below $60^{\circ}$, which is against
our preconceived idea about the orientation of the sample
members. However, as we have seen, this result could simply be a
consequence of the evolution of radio galaxies which undergo a
diminution of the power of their radio lobes power as they expand.


\subsection{The host galaxies}

\begin{figure}
\vspace{9.5cm}
\includegraphics{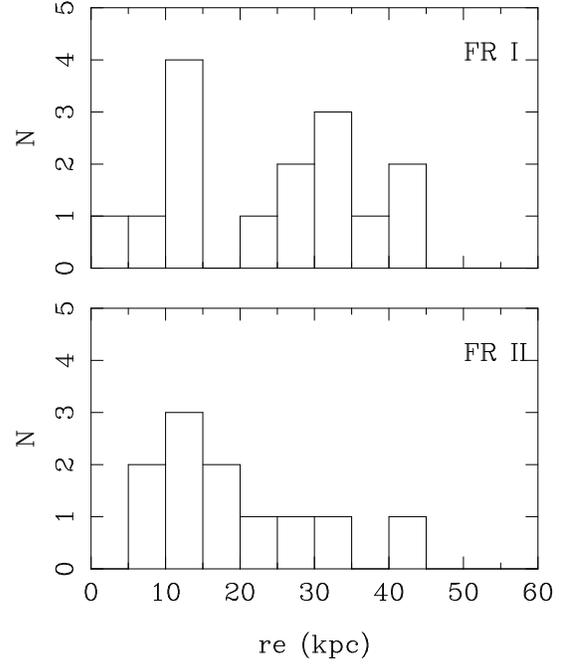} 
\caption{Distribution of the effective radius of the host galaxies
observed at optical wavelengths (Paper II)} 
\label{re_histo}
\end{figure}

As expected, the host galaxies of the sample
members for which a brightness profile could be derived (35\% of the
sample) are consistent with being of elliptical morphology (see Paper II). In
Fig.~\ref{re_histo} we show histograms of the effective radius
derived by fitting a $r^{1/4}$ profile to the brightness distribution,
separated for FR I and FR II type radio galaxies. Although the
statistics are rather poor, FR I radio galaxies tend to have larger 
effective radius than FR IIs, consistent with the result by Govoni et
al. (\cite{govoni}) obtained from a detailed study of a sample of low
redshift radio galaxies. These authors also find that the hosts of FR I radio
galaxies are more luminous than the hosts of FR IIs (see also Owen \&
Laing \cite{owen}).

Laing \& Bridle (\cite{laing2}) present strong evidence that the
deceleration of the jets in the low luminosity radio galaxy
\object{3C\,31} is produced by a mechanism of entrainment of thermal
matter in a dense interstellar medium (ISM) across the jets boundary.  
If the ISM density is directly related to the observed luminosity
(i.e., more material coming from stellar mass losses), then the luminosity
difference in FR Is and FR IIs could be an indirect measure of the degree of
deceleration in the jets of these objects. Moreover, the higher
luminosity of FR Is with respect to FR IIs might be related to the
separation between the two types of radio galaxies being also
dependent on optical luminosity (Ledlow \& Owen
\cite{ledlow}). However, Govoni et al. (\cite{govoni}) find a very
large scatter in the radio - optical luminosity plane and,
unfortunately, we have no reliable data on optical luminosity for our
sample to make a similar study. But this idea contrasts with the
sharp FR I/II break we find in radio power, independently of optical
luminosity (Section \ref{fr12}).

If we consider all the galaxies in our sample with spectroscopic
information, we find that 65\% of the FR I type radio galaxies have
optical spectra characterized by absorption lines only. On the other
hand, 71\% of FR II type radio galaxies show emission lines in their
optical spectra. This result is consistent with the unified model of
radio sources (e.g. Antonucci et al. \cite{antonucci}), which relates
FR I radio galaxies with BL-Lacs (generally without strong emission
lines) and FR II radio galaxies with radio quasars (with prominent
emission lines). But it also brings up the question about the FR I -
FR II dichotomy, since this result connects the type of large scale
radio structure with the properties of the optical spectrum in the
active core of the galaxy.

The differences in the spectral properties between the two families of
radio galaxies do not correspond with different properties in the
parsec scale jets observed with VLBI, which present similar radio
structures and evidence of relativistic flows (e.g. Giovannini et
al. \cite{giov01}). However, radio observations at sub-kiloparsec
scales show that the differences between the two families of radio
galaxies might arise at distances of a few hundreds of parsecs from
the central core (e.g. Lara et al. \cite{lara99}), a size coincident
with that estimated for the narrow line region in Seyfert galaxies
(Netzer \cite{netzer}).

Baum et al. (\cite{baum}) conclude, from the study of a sample of FR I
and FR II type radio galaxies, that while emission-line gas in FR IIs
is photoionized by nuclear UV continuum from the AGN, in FR Is the
emission line gas is most possibly energized by the host galaxy
itself. They argue as most plausible that there is a fundamental
difference in the central engines (accretion rate) and/or immediate
accretion region around the engine in FR I and FR II radio galaxies.
In a similar fashion, Ghisellini \& Celotti (\cite{ghisellini})
suggest that the FR I-FR II dichotomy is mainly controlled by the
properties of the accretion process, and even consider the possibility
of an FR II to FR I evolution of individual objects in some cases (see Section
\ref{lumf}). Chiaberge et al. (\cite{chiaberge}) observe a sample of
FR I radio galaxies with the HST and affirm that ``the innermost
structure of FR I radio galaxies differs in many crucial aspects from
that of the other classes of AGN; they lack the substantial BLR, tori
and thermal disc emission, which are usually associated with active
nuclei'', and suggest that in FR Is ``accretion might take place in a
low efficiency radiative regime''.

Arguments which try to explain the FR I-FR II dichotomy based
solely on differences of the central engine are not free from
difficulties: first, the similar properties between parsec scale jets
in FR I and FR II radio galaxies imply the same (or very similar) jet
production mechanism for both types of sources, and second, the
existence of FR I/II sources with large scale properties common with
FR I and FR II radio galaxies (Gopal-Krishna \& Wiita
\cite{gopal2}). We support the idea that the central kiloparsec
region surrounding 
the active core must also play, together with the central engine, a 
crucial role in the FR I-FR II
dichotomy. It is in this region where differences in the radio
structures of the two families of radio galaxies become evident and
where the emission lines in the optical spectra are produced. 
Moreover, the properties of this region must be closely linked with
the conditions in the region of accretion, which define the source of
ionization or the different core power.  
In a small fraction of sources, the external medium at larger scales
(possibly at the galactic halo) might determine key properties of the
large scale radio structure (FR I/II sources, or environment induced
asymmetry).

\subsection{The power - size diagram}
\label{power}

\begin{figure}
\vspace{7.5cm}
\includegraphics{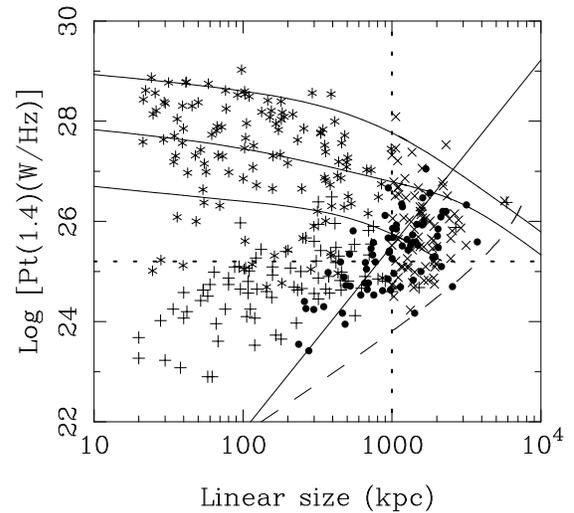} 
\caption{Power - size diagram of radio galaxies. Filled circles
correspond to our sample; crosses correspond to the B2 sample;
x-shaped crosses correspond to a compilation of GRGs (Ishwara-Chandra
\& Saikia \cite{ishwara}, Schoenmakers et al. \cite{arno1}, Machalski
et al. \cite{machalski}); asterisks correspond to the Peacock \& Wall
(\cite{peacock}) sample. The dashed line represents the sensitivity
limit of the NVSS for a 16$'$ extended 100 mJy source (see Paper
I). The horizontal dotted line marks the power break between FR I and
FR II radio sources. The vertical dotted line marks the definition of
GRGs. The curved continuous lines represent evolutionary tracks for
sources with jet powers of (top to bottom) $1.3\times 10^{40}$,
$1.3\times 10^{39}$ and $1.3\times 10^{38}$ W from Kaiser et
al. (\cite{kaiser}). The straight continuous line represents the trend
imposed by our selection criteria in flux density and source angular
size.} 
\label{pd}
\end{figure}

The relation between the intrinsic size and the radio power of radio
galaxies is a powerful tool for studying the evolution of radio
galaxies (Shklovskii \cite{shklovskii}, Scheuer \cite{scheuer}, Neeser
et al. \cite{neeser}, Kaiser et al. \cite{kaiser}). We display in
Fig.~\ref{pd} the power - size diagram for a large compilation of
radio galaxies taken from the B2 sample, the Peacock \& Wall
(\cite{peacock}) sample, our sample of large angular size radio
galaxies (Paper I), and a compilation of GRGs from Ishwara-Chandra \&
Saikia (\cite{ishwara}), Schoenmakers et al. (\cite{arno1}) and
Machalski et al. (\cite{machalski}). Both FR I and FR II radio
galaxies are represented in this plot. In order to separate the
influence of the sample selection criteria and of the evolution of
radio galaxies we have plotted also a line representing the
sensitivity limits affecting our sample, as discussed in Paper I. This
sensitivity limit mostly affects the selection of FR I radio galaxies,
and in principle of very large FR IIs. We also
plot the evolutionary tracks of FR II radio galaxies determined by
Kaiser et al. (\cite{kaiser}) from a model of the cocoon of FR II
radio galaxies which takes into account the energy loss processes for
the relativistic electrons. The lines plotted correspond to their
Fig.~1, for three different jet powers, and to the case when the
energy of the magnetic field and of the particles is in equipartition
(their case 3). This model explains the dearth of high luminosity
GRGs, even if recent systematic searches of GRGs are reaching lower
and lower sensitivity limits (Paper I; Schoenmakers et
al. \cite{arno1}; Machalski et al. \cite{machalski}). Moreover, the
energy losses in the radio lobes of extended radio galaxies might also
explain the relatively high core prominence in these objects 
(Sect.~\ref{pcore}).

Note that the apparent correlation between the radio power and the
source size in our sample (filled circles in Fig.~\ref{pd}) is well
reproduced taking into account the biases introduced by the sample
selection criteria in total flux density and angular size, thus the
need of considering several samples with different selection criteria 
to derive valuable conclusions from power-size relations.

The lack of high power GRGs is well visible in Fig.~\ref{pd} and confirmed by
the large sample of radio sources: in the linear size range $<$1
Mpc we have many sources with radio power between $10^{29}$ and
$10^{28}$ W/Hz, while very few GRGs have radio power
$>10^{27}$W/Hz. It cannot be justified by selection effects.  
A lack of giant radio sources is expected by evolutionary models of radio
sources (Kaiser et al. \cite{kaiser}; Blundell et
al. \cite{blundell}). We estimate that a radio
power loss of a factor 10 to 100 is required from the size--power relation.
This loss is in remarkably agreement the total radio power loss expected
from the $P_c$ vs $P_t$ correlation (see Sect.~\ref{pcore}).

\section{The giant radio galaxies}
\label{grg}

We summarize here the results presented in previous
sections, but from the point of view of giant radio galaxies. To
improve the statistics of this type of radio galaxies, we consider
when possible a compilation of 115 GRGs taken from Ishwara-Chandra \&
Saikia (\cite{ishwara}), Schoenmakers et al. (\cite{arno1}), Machalski et
al. (\cite{machalski}), and our sample (Paper I) (the same compilation
shown in Fig.~\ref{pd}).

From Fig.~\ref{pd} we find that most GRGs have sizes below 3 Mpc, with
only a limited number of objects surpassing this size. A cutoff in the
linear size of GRGs was already noted by Ishwara-Chandra \& Saikia
(\cite{ishwara}) from the analysis of a sample of 50 GRGs, and by
Schoenmakers et al. (\cite{arno2}) from a sample of 47 GRGs. Our
sample helps to confirm this cutoff between 2 and 3 Mpc, although we
note it
might be a combined effect of the decrease in luminosity with source
size (Kaiser et al. \cite{kaiser}), the sensitivity limitation of the
NVSS and the bias introduced by our selection criteria (see
Fig.~\ref{pd}). 

\begin{figure}
\vspace{6cm}
\includegraphics{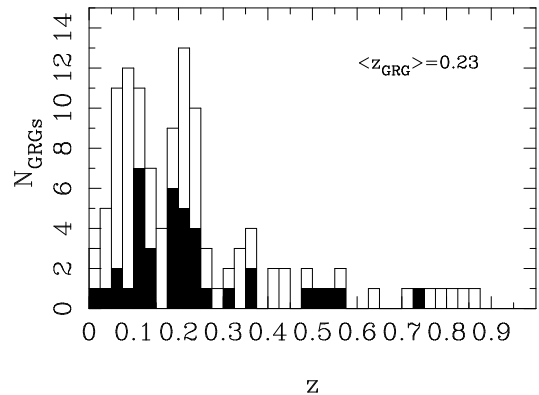} 
\caption{Distribution of redshift of giant radio galaxies. The white histogram
corresponds to a compilation of 115 GRGs (see text for
references). The black histogram corresponds to the GRGs in our
sample.} 
\label{lumsize}
\end{figure}

In Fig.~\ref{lumsize} we plot the distribution with redshift of the
GRGs in our sample, together with the same distribution for the
compilation of GRGs. We find that most known GRGs have a redshift
$z \le 0.25$. In our sample we know it is mostly due to our selection
criteria: we do not select sources smaller than 1 Mpc at $z>0.2$, and
the minimum selectable size increases rapidly with $z$ (see Paper
I). That means that at $z > 0.2$ many GRGs are being missed and only
those with sizes of about 2 Mpc or larger can be selected, but these
are rare objects.  Moreover, most ($\sim 77$\%) of the GRGs in our
sample are FR II type radio galaxies. This is not unexpected because
of the properties of FR I radio galaxies: first, they have lower power
than FR IIs, and second, their surface brightness decreases rapidly
with distance from the activity center, so sensitivity limitations
make difficult to observe their emission at very large distances from
the core.

If we compare the arm-length ratio, $r$, of the GRGs of FR II type in
our sample with that of the FR IIs in the whole sample, we do not find
any significant difference. GRGs have an average $r=0.79\pm 0.14$, fully
consistent with the value obtained for the whole sample ($r=0.78\pm
0.14$). In consequence, we do not find evidence that GRGs are more
asymmetric than smaller galaxies, in contrast with Schoenmakers et
al. (\cite{arno1}).

There are several circumstances one can think of which might help a
radio galaxy to become a GRG: {\em i)} a lower density external medium
so that the expansion of the radio source is not hampered, {\em ii)}
an older age to allow the radio galaxy to expand over large distances,
or {\em iii)} a higher core power to provide the jets the necessary
thrust to reach Mpc scales. Our sample allows us to compare the core
power of GRGs and smaller radio galaxies. Fig.~\ref{D_Pc5} shows the
apparent correlation between core radio power and source size. Since
this correlation can be fully explained by the biases introduced in
the sample selection through the flux density and the angular size
limits (Sect.~\ref{pcore}), it is not possible to conclude that GRGs have
more powerful cores than smaller radio galaxies. On the other hand,
spectral aging analysis suggest that GRGs are
the older population of normal FR I and FR II radio galaxies (Mack et al. 
\cite{mack}, Schoenmakers et al. \cite{arno1}). But time alone is not 
sufficient if the external medium is very
dense. We suggest that it
must be 
the combination of two ingredients, an old age and a low density
environment, which results in a giant radio galaxy. 

\section{Cycles of activity in radio galaxies}

If the activity of a radio-loud AGN is the result of accretion of
matter onto a compact massive object, likely a black hole, then it is
reasonable to assume that the life of a radio source is determined by
the accretion rate. It should be then expected to find in samples of
radio galaxies a significant number of sources with signs of having
passed through periods of different degrees of activity produced by
possible changes in the accretion rate.

We detect a compact radio core in 100\% of the radio galaxies in our
sample, a fact that can be interpreted as core activity 
currently present in all selected objects;  there are no
switched-off cores, with perhaps the exception of \object{J2111+630}, a large
angular size FR II type radio galaxy with a very weak core and no
evident hotspots in its radio structure (see Paper I, but
unfortunately its redshift could not be determined so the radio power
could not be evaluated).

In our sample we do find objects, with active cores, but that present signs
of different phases, or cycles, of activity: \object{J0317+769} (FR I),
\object{J1745+712} (FR II), \object{J1835+620} (FR II),
\object{J2035+680} (FR I) and \object{J2340+621} 
(FR I) (see Paper I for images of the radio sources):
\begin{itemize}
\item J0317+769 presents a small FR I type morphology, with well defined 
lobes, and a long and diffuse tail which could be the result of an older 
period of activity.
\item J1745+712 shows a FR II type structure without evident hotspots,
and a small bright and symmetric ejection which might be the consequence 
of an enhancement of the activity.
\item J1835+620 has two symmetric bright components within a typical FR II 
structure (Lara et al. \cite{lara}).
\item J2035+680 is a FR I radio source with a very bright knot of emission 
in one of the jets, and diffuse extended, probably older, emission beyond.
\item J2340+621 presents a typical FR I radio structure, with symmetric and 
bright knots of emission which might be the consequence of an enhancement of 
the core activity.
\end{itemize} 
We note that reactivation, as we consider it here, is a more general
phenomenon than the ``double-double'' morphology described by
Schoenmakers et al. (\cite{arno3}), which can be considered as a
particular case within which only the radio source in our sample
J1835+620 properly fits.

The physical conditions under which a radio source can experiment a
process of reactivation are not known. 
It is believed that interaction and merging with neighboring galaxies 
can be critical for this, providing an efficient mechanism to remove
angular momentum and directing gas towards the active center of the
galaxy (Bahcall et al. \cite{bahcall}). However, 
we only find nearby companions in two of the previously mentioned galaxies
(J0317+769 and J1835+620, see Paper II). A much more detailed optical study of
the host galaxies with high angular resolution is
necessary to reach definite conclusions about the relation between
merging and reactivation of the radio emission.

\section{FR II to FR I evolution?}
\label{lumf}

In this section we address the question raised by several authors
(e.g. Baum et al. \cite{baum}, Ghisellini \& Celotti
\cite{ghisellini}) about the possibility of an FR II type radio galaxy
evolving into an FR I type source. For that, 
we compare the predictions derived from the radio
luminosity function (RLF) with our observations. Given that our sample
consists of large size objects, and for that reason probably older
than objects in other samples (from which the RLF is determined), we
expect that if such evolution exists it will manifest itself in an
evolved sample like ours, in the sense of having an over-population of
FR Is and fewer FR IIs than expected from the RLF.

We have used the Dunlop \& Peacock (\cite{dunlop1}) RLF at 2.7 GHz.  To
choose one among the several models presented by these authors is
irrelevant at the redshifts involved in our sample, since all models
are well constrained by the local RLF. We have used their free-form
model 5, which provides the best results for the LBDS Hercules sample
of mJy radio sources (Waddington et al. \cite{waddington}). In the
RLF, we have introduced the area of our survey and the flux density
limit of the sample transformed to 2.7 GHz assuming a spectral index
$\alpha = -0.75$. We have also compared the results with the recent RLF
determined by Willott et al. (\cite{willott}) at 151 MHz. As expected
for radio sources with redshift below 1, both RLFs provide similar
results.

\begin{figure}
\vspace{6cm}
\includegraphics{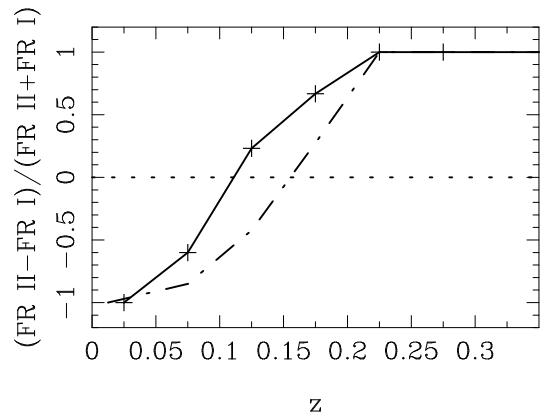} 
\caption{Relative number of FR I and FR II type radio galaxies as a
function of redshift. The continuous line represents the result
obtained from our sample. The dashed line is the result obtained from
the Dunlop \& Peacock (\cite{dunlop1}) RLF.} 
\label{abund}
\end{figure}

We define the parameter $x$ as the difference in the number of FR II
and FR I radio sources divided by the total number of radio sources
per redshift interval of 0.05. This parameter illustrates the relative
abundance of FR Is and FR IIs in the sample, and how this relative
abundance varies with the redshift. If all sources are of FR II type,
then $x=1$; if there are only FR I type radio galaxies, then $x=-1$. In
Fig.~\ref{abund} we display the parameter $x$ in the redshift range 0
to 0.35. At higher redshifts $x=1$, since we do not detect FR Is.
From the RLF we have computed, for each redshift interval, the number
of sources with radio power below and above the break power that
separates neatly the two families of radio galaxies
(Section~\ref{fr12}), and obtained the value of $x$ and its dependence
with redshift. It is not possible to introduce in the RLF the angular
size limit of our sample, so we assume that $x$ does not depend on the
source angular size (see note of caution below).  

We obtain that the RLF predicts well the lack of FR IIs at low
redshift, and the absence of FR I type radio galaxies at redshifts
above 0.25. However, from our data we find a larger number of FR II
type radio galaxies than predicted by the RLF. This result is not
consistent with an evolution of individual radio galaxies from FR II
to FR I type since in that case we should obtain the opposite result
in an evolved sample like ours: a larger abundance of FR Is. This
result, that is, the lack of evidence of a FR II to FR I evolution,
might be influenced by at least two factors: first, it is not possible
to discard a dependence of the parameter $x$ with the radio source
size, which might be partially masking the conclusions; second, but
related, the overabundance of FR II radio galaxies might be the result
of a bias in the selection of FR I radio galaxies due to sensitivity
limitation in the detection of extended emission. 

\begin{figure}
\vspace{14.5cm}
\includegraphics{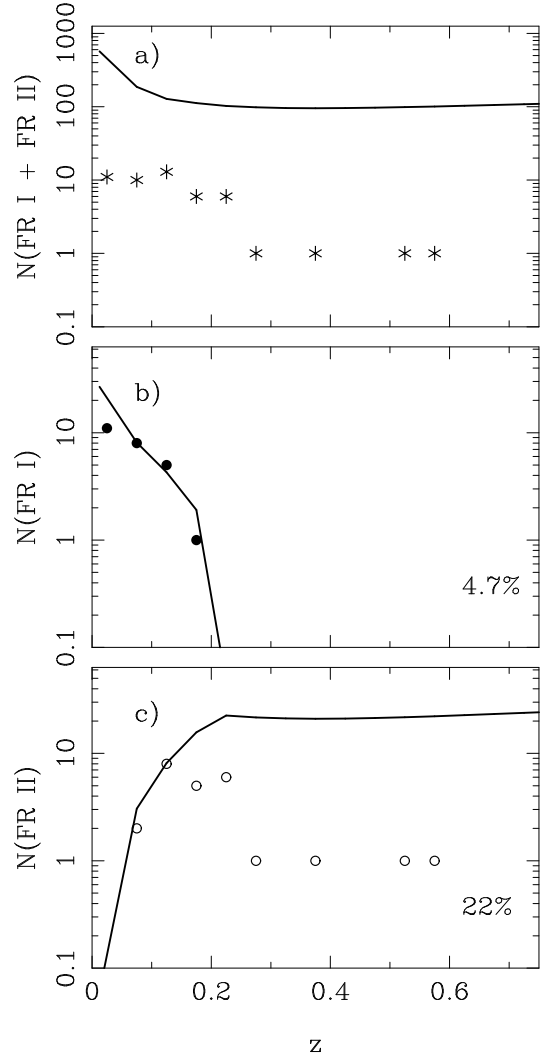}
\caption{{\bf a} Number of radio sources with flux density at 1.4 GHz above 
100 mJy and angular size above 4$'$ per redshift bin of 0.05 obtained
from our sample (asterisks). The continuous line represents the number
of sources with flux density above 100 mJy at 1.4 GHz and
unconstrained angular size obtained from the Dunlop \& Peacock
RLF. {\bf b} Filled dots represent the same as asterisks in {\bf (a)},
but only considering FR I type radio galaxies. The continuous line
represents the number of FR I type radio galaxies from the RLF, scaled
by a factor of 0.047. {\bf c} Same as {\bf (b)}, but for FR II type
radio galaxies. The continuous line results from the RLF after
applying a scaling factor of 0.22.}
\label{rlf}
\end{figure}

In Fig.~\ref{rlf} we display the number of radio galaxies per redshift
bin of 0.05, and compare the observational results with the
predictions of the RLF. As mentioned previously, the RLF cannot deal with an
angular size limit of $4'$ but we try here to estimate which is the
fraction of radio galaxies with angular sizes above this value,
comparing the RLF with our observations. In
Fig.~\ref{rlf}b, we show the number of FR I radio galaxies per
redshift bin. We find that the RLF, restricted to FR Is and scaled by
a factor of 0.047 predicts well the observed number of FR I objects at
a redshift $z\sim 0.1$, where we do not expect to have missed a
significant number of large sources in our sample due to the lack of
sensitivity to very extended emission (Paper I, Fig.6). Similarly, in
Fig.~\ref{rlf}c, we present our observed results for FR II radio
galaxies, and how the RLF, restricted to FR II radio galaxies and
scaled by a factor of 0.22 fits the observations at the same redshift range.
If we take into account the degree of spectroscopic completeness of
our sample (80\%; paper II) and assume that this completeness
distributes uniformly at all redshift bins (which is probably not
correct), the above factors change to 0.059 ($\sim 6\%$) and 0.28
($\sim 28\%$) for FR Is and FR
IIs, respectively. 
At redshifts higher than $z=0.1$, the RLF over-predicts the number of FR
IIs, which is not unexpected considering that, for example, at $z=0.4$
an angular size greater than 4$'$ implies a linear size larger than
1.5 Mpc, and such large sources are very rare (Section~\ref{grg}).

We can conclude that about 28\% and 6\% of the FR II and FR I sources,
respectively, at a redshift of about 0.1 and with a flux density above
100 mJy ($\log P_t(1.4) \ge 24.65$) have angular sizes larger than
$4'$. Considering Fig.~\ref{lum_funct}a, we find that all FR IIs and
about half of the FR Is are above this power level. Therefore, 28\% of
all FR IIs and about 3\% of all FR Is at $z\sim 0.1$ have angular
sizes larger than $4'$. 
Considering only FR II type radio galaxies whose physical size
can be well determined from radio observations, that means that
about one out of four FR II sources have linear sizes larger than 500 kpc,
that is are in an advanced stage of evolution. For example, assuming
an expansion velocity of 0.1c, one out of four FR II radio
galaxies would be older than $\sim 10$ Myrs. For typical GRG ages of a
few tens of Myrs, we infer from our analysis that a large fraction of
FR II radio galaxies (about 70\%) are younger than this and that radio
loud activity is a short-lived process in AGNs.

\section{Conclusions}

We present in this paper, the last of a series of three, the
general properties of a sample of 84 radio galaxies selected from
the NVSS, with total flux density at 1.4 GHz $\ge 100$ mJy,
declination above $+60^{\circ}$ and angular size larger than $4'$. This 
study is based on radio (Paper I) and optical (Paper II) observations of 
the sample members. The main conclusions of our work can be summarized as 
follows:
\begin{itemize}

\item From a pure morphological distinction of the sample members
according to its FR I or FR II type morphology, we confirm that both
groups are separated by a break in the total radio luminosity,
at $\log P_{t}$(1.4 GHz)$= 25.2$ (W/Hz).

\item There is a small population of radio sources (7\% in our sample)
which show simultaneously properties of FR I and FR II radio galaxies,
with a mean total radio power in the region of transition between the
FR I and FR II families.  


\item We find that asymmetry in the structure of FR I and FR II radio
galaxies can be explained as due to anisotropies in the medium through
which the jet propagates. Arguments based on orientation effects would
require too high jet-head advance velocities.

\item FR I radio galaxies have in general lower power cores than FR II
radio galaxies, although the distinction is not as clear as in the total
radio power. There is a correlation, independent of the FR I - FR II
dichotomy, between the core and the total radio power, consistent with
that derived by Giovannini et al. (\cite{giov88}, \cite{giov01}).
The prominence of the radio cores in our sample (evidenced through the
$P_{CN}$ orientation indicator) can be explained as a
result of the evolution of extended radio galaxies. The evolution of
extended radio galaxies implies a total power decrease in GRGs of a
factor 10--100, in agreement with Kaiser et al.(\cite{kaiser}).

\item Considering the known properties of FR I and FR II radio galaxies at 
parsec, subkiloparsec and megaparsec scales, and the properties of
their optical spectra, we suggest that the central kiloparsec region
plays a crucial role in explaining the FR I - FR II dichotomy, where
differences in the radio structure between the two families seem to
appear and whose properties might determine the presence or not of
emission lines. We note however that the properties of this region are 
surely linked with the properties of the accretion disk which ultimately 
determines the source of ionization and the different core radio power in FR I
and FR II sources, but apparently not the jet production mechanism.

\item The luminosity tracks predicted by the model of Kaiser et al. 
(\cite{kaiser}) explains well the dearth of high luminosity and large
size radio galaxies. Our study and other recent searches of GRGs, even
if much more sensitive and systematic than previous works on large
size radio galaxies, do not find high luminosity giant sources. 

\item We do not find evidence of the GRG phenomenon as being due to
stronger core radio power. Our observations are consistent with GRGs
as the older population of normal FR I and FR II radio galaxies
expanding in low density environments.

\item A compact radio core is detected in 100\% of the sample members,
indicating that core activity is present in all objects. However, we
find in our sample 5 objects with signs at kpc scales of having passed through
different phases of activity. Nearby galaxies are found in two of these
``reactivated'' objects, but more detailed observations are required
to study the role of galaxy merging in triggering different cycles of
activity.

\item Comparing the FR I and FR II abundances as a function of redshift with 
the predictions of the RLF, we do not find evidence of a possible FR II to 
FR I evolution of radio galaxies. 

\item From the RLF and our sample, we find that about one out of
four FR II type radio galaxies with $z\sim 0.1$ have linear sizes
larger than 500 kpc, thus being in an advanced stage of evolution.

\end{itemize}

\begin{acknowledgements}

This research is supported in part by the Spanish
DGICYT (grants AYA2001-2147-C02-01). LF and GG
acknowledges the Italian Ministry for University and Research (MURST)
for financial support under grant Cofin 2001-02-8773. The National Radio
Astronomy Observatory is a facility of the National Science Foundation
operated under cooperative agreement by Associated Universities, Inc. 

\end{acknowledgements}


\begin{thebibliography}{Normandin \& Kronberg 1980}

\bibitem[1987]{alexander} Alexander, P. \& Leahy, J. P. 1987, \mn,
225, 1

\bibitem[1993]{antonucci} Antonucci, R. 1993, \rev, 31, 473

\bibitem[1997]{bahcall} Bahcall, J.N., Kirhakos, S., Saxe, D.H.,
Schneider, D.P. 1997, \apj, 450, 486 

\bibitem[1995]{baum} Baum, S. A., Zirbel, E.,L., O'Dea, C. P. 1995,
\apj, 451, 88 

\bibitem[1974]{blandford} Blanford, R. D. \& Rees, M. J. 1974, \mn, 169, 395

\bibitem[1999]{blundell} Blundell, K. M., Rawlings, S. \& Willott,
C. J. 1999, \anj, 117, 677

\bibitem[2002]{cavaliere} Cavaliere, A. \& D'Elia, V. 2002, \apj,
571, 226

\bibitem[2000]{chiaberge} Chiaberge, M., Capetti, A., Celotti, A. 2000, \aa, 355, 873

\bibitem[1998]{nvss} Condon, J. J., Cotton, W. D., Greisen, E. W., et al. 1998, \anj, 115, 1693

\bibitem[1990]{deruiter} de Ruiter, H. R., Parma, P., Fanti, C., Fanti, R. 1990, \aa, 227, 351

\bibitem[1990]{dunlop1} Dunlop, J. S. \& Peacock, J. A. 1990, \mn, 247, 19

\bibitem[2003]{dunlop2} Dunlop, J. S., McLure, R. J., Kukula, M., et
al. 2003, \mn, 340, 1095

\bibitem[1974]{fanaroff} Fanaroff, B. L. \& Riley, J. M. 1974, \mn, 167, 31

\bibitem[2003]{georga} Georganopoulos, M. \& Kazanas, D. 2003, \apj,
589, L5

\bibitem[2001]{ghisellini} Ghisellini, G. \& Celotti, A. 2001, \aa,
379, L1

\bibitem[1988]{giov88} Giovannini, G., Feretti, L., Gregorini, L., Parma, P. 1988, \aa, 199, 73

\bibitem[1994]{giov94} Giovannini, G., Feretti, L., Venturi, T. et
al. 1994, \apj, 435, 116

\bibitem[2001]{giov01} Giovannini, G., Cotton, W. D., Feretti, L. et al. 2001, \apj, 552, 508 

\bibitem[1989]{gopal1} Gopal-Krishna, Wiita, P. J., Saripalli, L. 1989, \mn, 239, 173

\bibitem[2001]{gopal2} Gopal-Krishna, Wiita, P. J. 2001, \aa, 373, 100

\bibitem[2000]{govoni} Govoni, F., Falomo, R., Fasano, G., Scarpa, R. 2000, \aa, 353, 507

\bibitem[1999]{ishwara} Ishwara-Chandra, C. H. \& Saikia, D. J. 1999, \mn, 309, 100

\bibitem[1997]{kaiser} Kaiser, C. R., Dennett-Thorpe, J., Alexander, P. 1997, \mn, 292, 723

\bibitem[1994]{komissarov} Komissarov, S. S. \& Gubanov, A. G. 1994, \aa, 285, 27

\bibitem[2001]{kormendy} Kormendy, J. \& Gebhardt, K. 2001, in The
20th Texas Symposium on Relativistic Astrophysics, ed. H. Martel \&
J.C. Wheeler, AIP Vol. 586, 363

\bibitem[2002]{laing1} Laing, R. A. \& Bridle, A. H. 2002, \mn, 336, 328

\bibitem[2002]{laing2} Laing, R. A. \& Bridle, A. H. 2002, \mn, 336, 1161

\bibitem[1999a]{lara} Lara, L., M\'arquez, I., Cotton, W. D., et al. 1999a, \aa, 348, 699

\bibitem[1999b]{lara99} Lara, L., Feretti, L., Giovannini, et al. 1999b, \apj, 513, 197

\bibitem[2001a]{paperI} Lara, L., Cotton, W. D., Feretti, L., et al. 2001a, \aa, 370, 409 ({\bf Paper I})

\bibitem[2001b]{paperII} Lara, L., M\'arquez, I., Cotton, W. D., et
al. 2001b, \aa, 378, 826 ({\bf Paper II})

\bibitem[1996]{ledlow} Ledlow M.J. \& Owen F.N., 1996, \anj, 112, 9

\bibitem[1979]{longair} Longair, M. S. \& Riley, J. M. 1979, \mn, 188,
625

\bibitem[1999]{macchetto} Macchetto, F. D. 1999, \apss, 269/270, 269 

\bibitem[2001]{machalski} Machalski, J., Jamrozy, M., Zola, S. 2001, \aa, 371, 445

\bibitem[1998]{mack} Mack, K. -H., Klein, U., O'Dea, C. P., et al. 1998, \aa, 329, 421

\bibitem[2001]{mclure} McLure, R. J. \& Dunlop, J. S. 2001, \mn, 327, 199

\bibitem[1993]{morganti} Morganti, R., Killeen, N.E.B., Tadhunter,
C.N. 1993, \mn, 263, 1023

\bibitem[1995]{neeser} Neeser, M. J., Eales, S. A., Law-Green, J. D., et al. 1995, \apj, 451, 76

\bibitem[1990]{netzer} Netzer, H. 1990, in Active Galactic Nuclei, eds. T.J.L. Courvoisier \& M. Mayor, Springer-Verlag, 57

\bibitem[1989]{owen} Owen, F. N. \& Laing, R. A. 1989, \mn, 238, 357

\bibitem[1998]{owsianik} Owsianik, I. \& Conway, J. E. 1998, \aa, 337, 69

\bibitem[1981]{peacock} Peacock, J. A. \& Wall, J. V. 1981, \mn, 194, 331

\bibitem[1994]{readhead} Readhead, A. C. S., Xu, W., Pearson, T. J., et al. 1994, AAS Meeting, 182, 5307 

\bibitem[1978]{rees1} Rees, M. J. 1978, \nat, 275, 516

\bibitem[1974]{scheuer} Scheuer, P. A. G. 1974, \mn, 166, 513

\bibitem[1995]{scheuer2} Scheuer, P. A. G. 1995, \mn, 277, 331

\bibitem[2000a]{arno1} Schoenmakers, A. P., Mack, K. -H., de Bruyn, A. G., et al. 2000a, \aas, 146, 293

\bibitem[2000b]{arno3} Schoenmakers, A. P., de Bruyn, A. G., R\"ottgering, H. J. A., et al. 2000b, \mn, 315, 371

\bibitem[2001]{arno2} Schoenmakers, A. P., de Bruyn, A. G., R\"ottgering H. J. A., van der Laan H. 2001, \aa, 374, 861

\bibitem[1963]{shklovskii} Shklovskii, I. S. 1963, SvA, 6, 465

\bibitem[2001]{slee} Slee, O. B., Roy, A. L., Murgia, et al. 2001,
\anj, 122, 1172

\bibitem[1995]{urry} Urry, C. M. \& Padovani, P. 1995, \pasp, 107, 803

\bibitem[2001]{waddington} Waddington, I., Dunlop, J. S., Peacock,
J. A., Windhorst, R. A. 2001, \mn, 328, 882

\bibitem[1994]{wilkinson} Wilkinson, P. N., Polatidis, A. G., Readhead, A. C. S., et al. 1994, \apj, 432, L87

\bibitem[2001]{willott} Willott, C. J., Rawlings, S., Blundell, K. M., et al. 2001, \mn, 322, 536 

\bibitem[1995]{wilson} Wilson, A. S. \& Colbert A. J. M. 1995, \apj,
438, 62

\end{thebibliography}
\end{document}